\documentclass[conference]{IEEEtran}
\IEEEoverridecommandlockouts
\usepackage{amsmath,amsfonts}
\usepackage{array}
\usepackage{textcomp}
\usepackage{stfloats}
\usepackage{url}
\usepackage{verbatim}
\usepackage{graphicx}
\usepackage{color}
\usepackage{svg}
\usepackage{pdfpages}
\usepackage{booktabs}
\usepackage[linesnumbered,ruled,vlined]{algorithm2e}
\RequirePackage{letltxmacro}
\LetLtxMacro{\LaTeXtextbf}{\textbf}
\LetLtxMacro{\textbf}{\LaTeXtextbf}
\usepackage{cite}
\usepackage{amsmath,amssymb,amsfonts}
\usepackage{graphicx}
\usepackage{textcomp}
\usepackage{caption}
\usepackage{subcaption}
\usepackage[bookmarks=false]{hyperref} 
\usepackage{algpseudocode}
\usepackage{array}
\usepackage{enumitem,lipsum}
\setlist[itemize,enumerate]{leftmargin=*}
\usepackage{etoolbox}
\newtheorem{definition}{Definition}[section] 

\def\BibTeX{{\rm B\kern-.05em{\sc i\kern-.025em b}\kern-.08em
    T\kern-.1667em\lower.7ex\hbox{E}\kern-.125emX}}
\usepackage{tcolorbox}
\tcbuselibrary{breakable}

\makeatletter
\newcommand{\stackover}{\genfrac{.}{.}\z@{}}
\makeatother

\begin{document}

\title{EnerSwap: Large-Scale, Privacy-First Automated Market Maker for V2G Energy Trading}



\author{\IEEEauthorblockN{Ahmed Mounsf Rafik Bendada}
\IEEEauthorblockA{\textit{L3i Laboratory} \\
\textit{La Rochelle university}\\
La Rochelle, France \\
ahmed.bendada@univ-lr.fr}
\and
\IEEEauthorblockN{Yacine Ghamri-Doudane}
\IEEEauthorblockA{\textit{L3i Laboratory} \\
\textit{La Rochelle university}\\
La Rochelle, France \\
yacine.ghamri@univ-lr.fr}
}

\maketitle

\begin{abstract}
With the rapid growth of Electric Vehicle (EV) technology, EVs are destined to shape the future of transportation. The large number of EVs facilitates the development of the emerging vehicle-to-grid (V2G) technology, which realizes bidirectional energy exchanges between EVs and the power grid. This has led to the setting up of electricity markets that are usually confined to a small geographical location, often with a small number of participants. Usually, these markets are manipulated by intermediaries responsible for collecting bids from prosumers, determining the market-clearing price, incorporating grid constraints, and accounting for network losses. While centralized models can be highly efficient, they grant excessive power to the intermediary by allowing them to gain exclusive access to prosumers’ price preferences. This opens the door to potential market manipulation and raises significant privacy concerns for users, such as the location of energy providers. This lack of protection exposes users to potential risks, as untrustworthy servers and malicious adversaries can exploit this information to infer trading activities and real identities. This work proposes a secure, decentralized exchange market built on blockchain technology, utilizing a privacy-preserving Automated Market Maker (AMM) model to offer open and fair, and equal access to traders, and mitigates the most common trading-manipulation attacks. Additionally, it incorporates a scalable architecture based on geographical dynamic sharding, allowing for efficient resource allocation and improved performance as the market grows.

\end{abstract}


\maketitle
\begin{tcolorbox}[breakable,boxrule=1pt,colframe=black,colback=white]
\scriptsize Paper accepted at the 27th International Conference on Modeling, Analysis and Simulation of Wireless and Mobile Systems (MSWiM)
\end{tcolorbox}

\section{Introduction}\label{sec:introduction}

\IEEEPARstart{T}hroughout history, people have traded valuable goods to meet their needs, from basic necessities like food and tools to precious commodities like gold. In the last decades, fossil fuels such as oil and natural gas have been the backbone of global trade, but this trade was strictly one-way, moving from producers to consumers with no return flow. However, with the growing focus on reducing carbon emissions, there is increasing interest in renewable energy sources and electric vehicles (EVs). Unlike fossil fuels, these technologies enable a two-way exchange, where electricity can not only be consumed but also returned to the grid \cite{surv1}. This shift has paved the way for the development of electricity markets, positioning electricity itself as a tradable asset in the evolving energy landscape.

As this transition unfolds, EVs are emerging as key players in the future of transportation and energy management. EVs are not only reducing dependence on fossil fuels but also transforming the way we interact with electricity. With the advent of vehicle-to-grid (V2G) technology, EVs can engage in bidirectional energy exchanges with the power grid, serving as both consumers and suppliers of electricity \cite{v2g}. This capability has led to the development of local electricity markets, where EV owners, equipped with the ability to store and return energy, become prosumers, both producing and consuming electricity \cite{V2G2022integration}.

These electricity markets are typically confined to small geographic areas with a limited number of participants, allowing intermediaries, especially energy providers, to dominate the market, which brings several key challenges. One issue is the potential for price manipulation, where rates are imposed on all participants without transparency \cite{8972609}. Another concern is the centralized control of sensitive information, which creates serious privacy risks for traders, as their location and transaction data could be exposed to malicious actors. Furthermore, this centralization increases the vulnerability of users to cybersecurity threats, where adversaries may launch sophisticated attacks to exploit personal and trading information \cite{AMM}.

The limitations of centralized systems highlight the need for a more secure and transparent approach to energy trading. Decentralization provides a solution by eliminating intermediaries and distributing control more equitably across the market. Decentralized exchanges (DEXs) \cite{DEX-ACM}, initially popularized in the cryptocurrency space, are built on blockchain technology and enable direct peer-to-peer transactions without the need for a central authority. In crypto, DEXs have empowered users by allowing them to trade assets directly from their wallets using smart contracts, ensuring transparency and reducing manipulation. 

However, DEXs come with their own set of challenges. They are susceptible to various security threats \cite{DEX-threats}, such as sandwich and front running attacks \cite{sandwich, dex-mev}, where malicious actors manipulate trade prices to profit from user transactions; liquidation attacks \cite{liquidation}, which take advantage of leveraged positions to trigger forced liquidations at unfavorable prices; and arbitrage attacks, where attackers exploit price differences between markets for personal gain \cite{dex-mev}. Additionally, smaller markets often struggle with a lack of liquidity, which can further hinder smooth trading and price stability.

Driven by the challenges outlined above, this paper offers the following key contributions:

\begin{itemize}
    \item We present a decentralized trading model tailored for small markets that does not require any intermediary or third party, leveraging Automated Market Makers (AMM) to enhance efficiency and market participation.
    \item We introduce how we apply a layer of privacy in the trading phase to mitigate the attacks launched by the untrusted peers of the networks by using succinct zero-knowledge proofs, commitment schemes, and Secure multiparty computation to ensure not only the privacy of the trades but also the well-formedness and efficiency of the various operations.
    \item We propose a publicly verifiable, trust-minimized framework that records all transactions on an append-only blockchain ledger, is rigorously proven secure and privacy-preserving, and is implemented on a scalable, dynamically geo-sharded blockchain architecture that adapts to different market sizes and use-case demands. The complete solution has been implemented as a proof of concept and is publicly available on GitHub \footnote{\href{https://github.com/0xmoncif213/EnerSwap}{https://github.com/0xmoncif213/EnerSwap}}.
\end{itemize}
The remaining organization of this paper is as follows. First, the existing
related literature is summarized in Section \ref{sec:SOTA}. The preliminaries are introduced in Section \ref{sec:prem}. Section \ref{sec:fram} presents the security properties and describes the overall framework proposed in this paper  . Section \ref{sec:bblocks} details the proposed cryptographic building blocks used in the framework . Section \ref{sec:security} formally analyzes our framework’s resilience to potential attacks. Finally, \ref{sec:eva} is devoted to the proof of concept implementation and its performance analysis, and Section \ref{sec:conclusion} concludes our paper and discusses future work.

\section{RELATED WORK}
\label{sec:SOTA}
\quad Existing decentralized energy-trading schemes fall into three principal categories: double-auction mechanisms, bilateral trading, and game-theoretic models; however, all struggle with scalability, privacy, and liquidity when applied to high-frequency EV charging. Double-auction designs \cite{doubleAuc} require continuous bid broadcasts until a clearing price emerges, which produces substantial on-chain traffic as numerous EVs connect within short intervals. Because bids are publicly stored, this also reveals individual charging patterns. Bilateral contracts \cite{bilateral} mitigate traffic by shifting negotiation off-chain, yet remove transparent price discovery and leave regulators without verifiable real-time information. Game-theoretic approaches range from seller-only markets \cite{47} to centralized Stackelberg configurations \cite{46} and decentralized Stackelberg or crowdsourcing variants implemented on blockchain \cite{48,80,70}; despite their formal elegance, equilibrium computation and state updates scale poorly with participant count, while published strategies or commitments expose household or fleet usage profiles. Several recent studies extend these paradigms but inherit similar limitations. Li et al. \cite{70} introduce an energy-coin and credit-based payment scheme on a consortium blockchain, complemented by Stackelberg pricing, yet they omit a formal double-spending analysis and do not present an implemented prototype. PriWatt \cite{76} combines smart contracts, multi-signatures, and anonymous messaging to secure transactions and hide identities, but offers no details on its mining process or reward structure, raising questions about throughput and incentive design. A consortium-blockchain proposal in \cite{77} replaces proof-of-work with proof-of-stake and adds cryptographic data protection and pseudonyms, although it lacks empirical evaluation of communication overhead and energy cost. Garg et al. \cite{78} couple blockchain with elliptic-curve–based hierarchical authentication for privacy in V2G trading, but the threat model is under-specified and the mechanism remains untested. Finally, a Hyperledger-Fabric–based crowdsourced energy-sharing framework \cite{80} automates EV charging, load deferral, and renewable integration, yet does not address attacks from malicious crowdsourcers or other stakeholders. Across these studies, fragmented orders mean that, at any instant, a buyer may not locate a compatible seller and vice versa; this lack of counterparties constitutes a liquidity shortage, an inability to trade promptly without affecting price. Collectively, current solutions advance the theory of decentralized energy markets but still fall short of the combined requirements for rapid, low-value EV transactions, rigorous privacy preservation, demonstrable scalability, and sufficient market liquidity.
\cite{AMM} introduced a concentrated-liquidity AMM for local energy trading that greatly deepens liquidity and widens participation. Yet the design still omits user-privacy safeguards and an explicit treatment of transaction fees omissions that could ultimately curb users' engagement. While \cite{icbc}addresses privacy by proposing an order-book-based, privacy-preserving marketplace, its solution targets generic goods rather than energy and offers no concrete deployment architecture.
\section{Background}
\label{sec:prem}
\subsection{Sharding on Blockchain}
Sharding was initially introduced in database systems to enhance both performance and availability. In the context of blockchain, sharding refers to splitting the network’s nodes into smaller groups, known as shards, where each shard is responsible for processing a subset of the network’s transactions.  There are multiple types of sharding: deterministic, probabilistic, and dynamic \cite{sharding}. Regardless of the type, sharding allows for parallel transaction processing, reduces the workload on individual nodes, and enhances throughput.
\subsection{AMMs model} Unlike centralized exchanges (CEXs), which rely on order books and custodial control of assets, AMMs use smart contracts to automate and facilitate trades. This allows users to maintain full control of their assets while interacting directly with the protocol. In an AMM-based market, traders interact with the Liquidity Pool (LP) instead of trading directly with each other. The LP funded by liquidity providers, holds two or more types of tokens that are exchanged on a blockchain. In our case, we consider two tokens: the E-token (which represents energy), and the M-token (which represents money). When a participant wants to purchase energy, they deposit M-tokens into the LP and receive E-tokens in return. Likewise, a participant selling energy deposits E-tokens into the LP and receives M-tokens. \\
 The basic formula for an AMM is expressed as \cite{adams2021uniswapv3}: 
 \begin{equation}
         E_{LP} .  M_{LP} =  C
    \label{eq:AMM}
\end{equation}
where, \\
$E_{LP} =$ Quantity of E-tokens in the LP \\
$M_{LP} =$ Quantity of M-tokens in the LP \\
$C =$ Constant AMM product \\

The current state of LP is described by:  ($E^{i}_{LP}$,$M^{i}_{LP}$) and the price of LP is given by:  \\
\begin{equation}
    p^{i}_{LP} = \dfrac{M^{i}_{LP}}{E^{i}_{LP}} = \dfrac{C}{(E^{i}_{LP})^2} = \dfrac{(M^{i}_{LP})^2}{C}
\end{equation}

Suppose the next transaction is a buy order of $E_{i+1}$ E-tokens. The number of E-tokens in the liquidity pool decreases by $E_{i+1}$. Since $C$ must remain constant, the corresponding amount of M-tokens in the liquidity pool can be calculated by Eq \ref{eq:AMM}. \\
The new state of LP will be: \\
\begin{equation}
    E^{i+1}_{LP} = E^{i}_{LP} - E_{i+1}  , 
    M^{i+1}_{LP} = \dfrac{C}{E^{i}_{LP} - E_{i+1}}
\end{equation}
The quantity of M-tokens that the buyer needs to provide for this transaction is: 
\begin{equation}
    M_{i+1} = M^{i+1}_{LP} - M^{i}_{LP} = \dfrac{C .  E_{i+1}}{E^{i}_{LP}(E^{i}_{LP} - E_{i+1})}
\end{equation}
The cost of energy for the buyer is: 
\begin{equation}
    p_{i+1}= \dfrac{M_{i+1}}{E_{i+1}} = \dfrac{C}{E^{i}_{LP}(E^{i}_{LP} - E_{i+1})}
\end{equation}
It is evident that $p_{i+1}$ changed from the initial LP price $p^{i}_{LP}$. This Delta is the price impact of the transaction can be calculated by:
\begin{equation}
    \Delta p_{i+1}= \dfrac{p_{i+1} - p^{i}_{LP} }{p^{i}_{LP}} = \dfrac{E_{i+1}}{E^{i}_{LP} - E_{i+1}}
\end{equation}
For a well-functioning market, the price variance should be as small as possible. This can be achieved when $E^{i}_{LP} >> E_{i+1}$. In other words, the quantity of the energy provided by liquidity providers should be much larger than the individual consumption size, which is the case for energy trading between EVs and Energy providers.
\section{proposed scheme}
In this section, we present $EnerSwap$, a large-scale geo-sharded AMM for peer-to-peer energy trading that fuses zk-SNARKs, Pedersen commitments, and secure multi-party computation to safeguard participants' data.  We first outline our assumptions and threat model, then describe the framework’s architecture and step-by-step workflow in detail.
\label{sec:fram}
\begin{figure}
    \centering
    \includegraphics[scale=0.22]{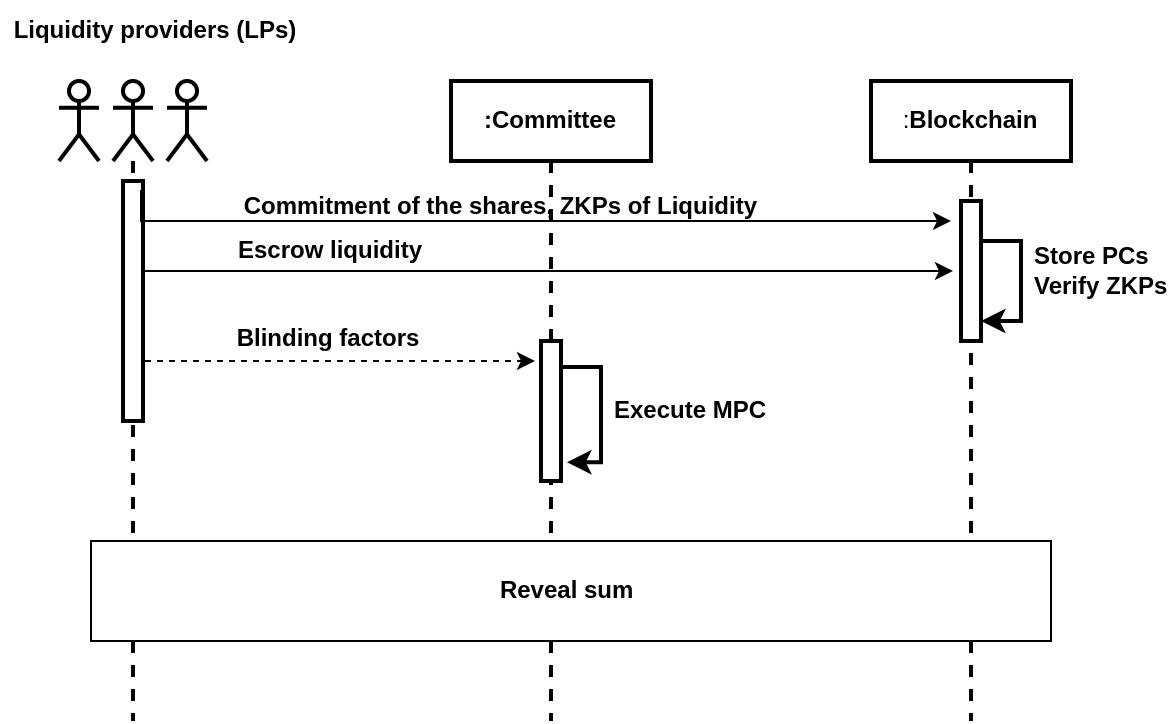}
    \caption{Init phase of the proposed framework}
    \label{fig:init}
\end{figure}

\subsection{Model and Assumptions}
\quad We rely on the following trust assumptions for the entities within our system. Our model consists of peers, who are regular users of a P2P network. Peers can be Traders (buyers and sellers) or Liquidity Providers (LPs) and interact with the platform via their real identities. We assume that all users are registered and uniquely identified by cryptographic keys, peers can exchange liquidity with the pool, and can demonstrate their balances to other peers using appropriate ledger transactions. The ledger keeps a record of all transactions that have happened in the network and can be verified by any third party.
\label{sec:assumptions}
\begin{itemize}
\item \textit{Traders:}
We assume that all traders are untrusted, with financial motivations to deceive about asset ownership, manipulate account balances, and try to exploit the system.
\item \textit{Blockchain nodes:}
While some nodes may be malicious, we rely on a two-thirds honest majority to uphold the integrity of the blockchain under Byzantine-tolerant consensus. Adversarial nodes may also collaborate with traders in attempts to undermine the system.
\item \textit{Committee Nodes}: We assume a dynamic permissioned model for selecting committee nodes \cite{MPC}.
\end{itemize}
\subsection{Threat Model}
\label{sec:security-prop}
\subsubsection{\textbf{Security proprities}}
We now define the security properties of our system. We have the marketplace operate in steps, with trades received from traders. The necessary security properties are listed below: \\
\textit{Confidentiality of Trades:}
Throughout the trading phase, a participant discloses nothing about the exact quantity they intend to buy or sell, only that their account balance exceeds the order rate $K$. This single piece of information is submitted through ZKP to the on-chain verification contract for validation. The actual bid amounts and exchange rates remain hidden until the trading process has concluded, at which point they are publicly revealed.\\
\textit{Confidentiality of traders' accounts:} The trader reveals no details of their full balance; they supply a ZKP that proves their balance exceeds a value $K (with 0 < K < balance)$, while $K$ itself is represented by shares embedded in a Pedersen commitment and serves as the exact amount locked into the trade. \\
\textit{Market Availability and integrity:} The marketplace is designed for high availability, with safeguards against Denial-of-Service and Blockchain-related attacks. The protocol enforces a buy rate that is always greater than or equal to the sell rate, ensuring that no buyer ever pays above their bid price and no seller ever receives less than the market instant price. 
\subsubsection{\textbf{Attacks (security and privacy)}}
\quad The primary objective of a trading system is to guarantee that all market transactions are fair and tamper-proof. Accordingly, beyond the core security and privacy requirements already outlined, the system must also withstand the following types of attacks: \\
\textit{Front running}: happens when a trader has advanced knowledge of a large, pending transaction, and then uses this information to place their own trade before the original one is executed. By doing this, they profit from the expected price movement caused by the large trade, gaining an unfair advantage over others in the market. \\
\textit{Sandwich:} is a type of front-running often seen in AMMs. In this attack, an attacker takes advantage of the price slippage that occurs during a large trade by placing two strategically timed trades, one before and one after the target transaction. This attack manipulates the price movement to the attacker's advantage while negatively affecting the victim’s trade. \\
\textit{Arbitrage:} is a trading strategy that involves taking advantage of price differences for the same asset in different markets or forms. The trader buys the asset at a lower price in one market and simultaneously sells it at a higher price in another, thereby profiting from the price discrepancy. \\
In addition to the above security properties. We present a formal security discussion in section \ref{sec:security}. A  more detailed analysis of trading attacks will be provided in the extended version of this work.
\subsection{Protocol description}
\begin{figure}
    \centering
    \includegraphics[scale=0.22]{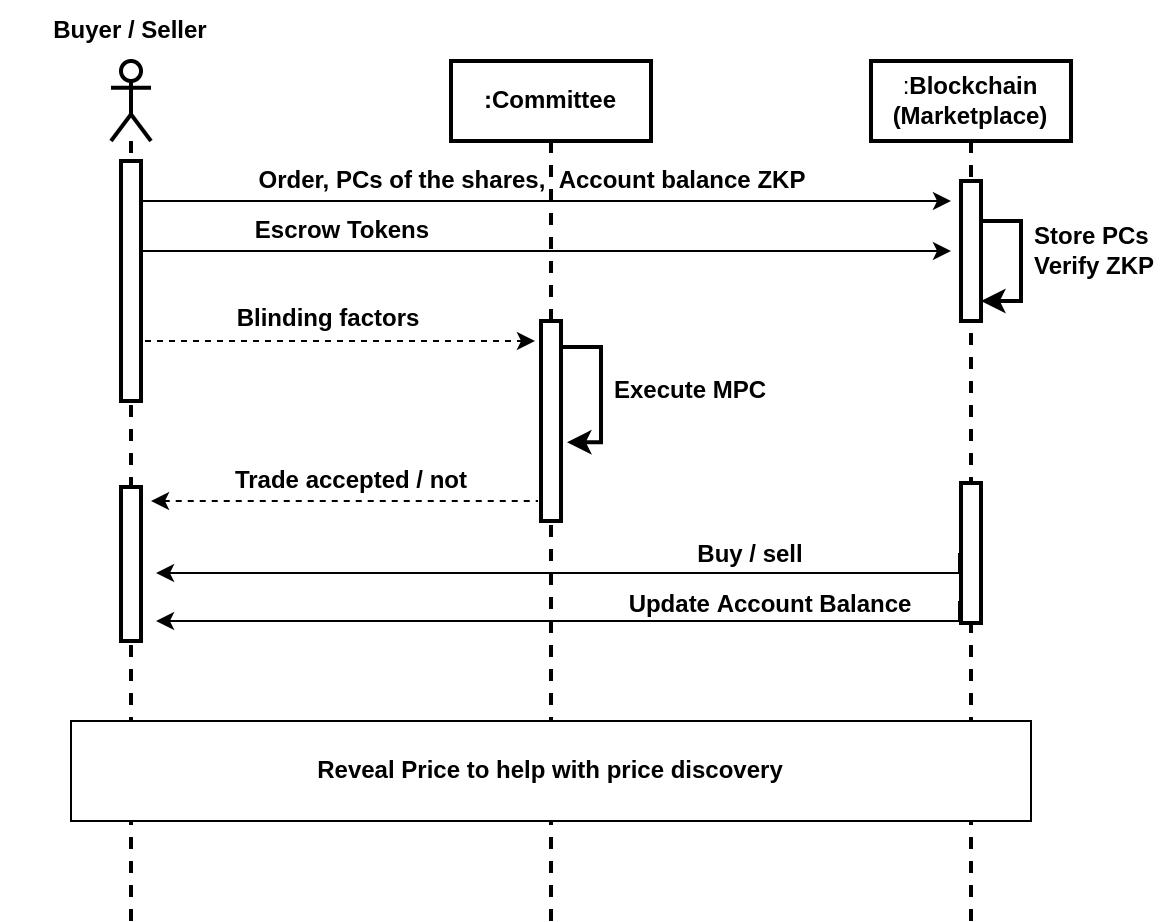}
    \caption{Workflow of the proposed framework}
    \label{fig:workflow}
\end{figure}
\begin{figure}
    \centering
    \includegraphics[scale=0.13]{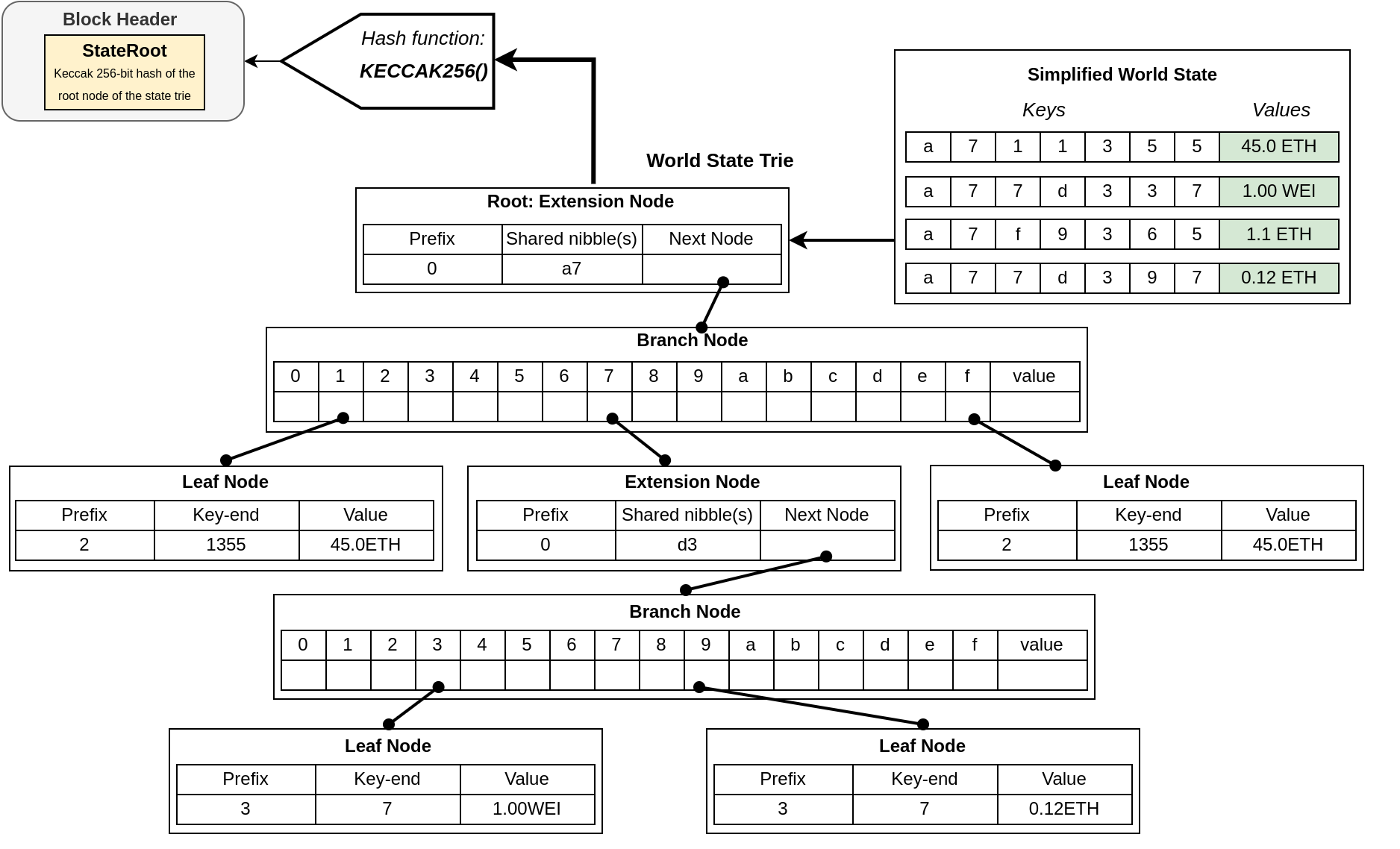}
    \caption{Simplified Ethereum Merkle Patricia Trie \cite{wood2025ethereum}}
    \label{fig:Patricia}
\end{figure}

\label{sec:protocol}
\textbf{1) Init phase:}
Our protocol begins with an \emph{initialization phase} in which each liquidity provider (LPs), as explained in \ref{fig:init} generates \emph{ Pedersen commitments} for the private liquidity they are willing to provide into the pool, \(x\). Formally, the LP selects a large prime \( p \), two generators \( g \) and \( h \) in a suitable cyclic group, and a random blinding factor \( r \). They compute
\begin{equation}
\label{eq:PC}
  \mathrm{Comm}(x, r) \;=\; g^{\,x} \,h^{\,r} \;\bmod\; p
\end{equation}

and store these commitments on-chain, alongside a \emph{zero-knowledge proof} (ZKP) certifying that the committed value \( x \) is non-negative and above a threshold \(k\) and owned by the LP.
As we use Ethereum as a blockchain, each account’s address is first $Keccak256$ hashed and split into 64 hexadecimal nibbles that steer a path through the \textit{Merkle Patricia Trie} (MPT), the leaf reached on that path stores the Recursive Length Prefix (RLP) encoded account tuple:
\begin{equation}
\{nonce, balance, storageRoot, codeHash\} 
\end{equation}
Every node’s RLP is hashed, and those hashes cascade upward to the single 32-byte $stateRoot$ recorded in the block header. The Patricia proof which is the ordered stack of encoded nodes along that path, obtainable via the function $eth \textunderscore getProof$ lets anyone recompute each node’s hash, follow the nibbles, decode the leaf, and confirm that the balance field is exactly what was committed, any alteration to the balance or path would flip a hash and break the match with the published $stateRoot$, so a successful verification cryptographically proves the precise balance without revealing other accounts.
Because the Merkle Patricia proof is inseparably bound to the StateRoot of one concrete block, the only way to pass verification is for the leaf to match the root that appears in that block header; a prover cannot cherry pick an older root without also revealing the block number (every header encodes its height, hash of the parent, timestamp). Figure \ref{fig:Patricia} illustrates how the MPT proofs are constructed. When a client asks a node for the latest finalized header and compares its StateRoot to the one referenced in the proof, two things happen:
\begin{itemize}
    \item \textit{{Freshness check:}} if the roots differ, the proof obviously isn’t for the present state; if the roots match, you know the balance comes from exactly that live block.
    \item \textit{{Unforgeability:}}
    altering a historic header to smuggle in a fake root would require rewriting that block’s hash, which would in turn break the hash-linked chain all the way to the tip, the attacker would have to re-win Ethereum consensus, since that block, an economically impossible task under Proof-of-Stake finality.
\end{itemize}
Practically, the verifier smart contract accepts a proof only if it comes from a block with at least N confirmations and a timestamp within an approved freshness window, ensuring the reported balance is the account’s current on-chain state, not a stale snapshot.
Because of the binding and hiding properties of Pedersen commitments, no information about the private balances leaks at this stage. The on-chain contract verifies the ZKP without learning \( x \) and stores the commitments. Simultaneously, each $LP_i$ privately transmits its blinding factors \( r_{ij} \) and shares of \( x_{ij} \) to a selected \emph{secure multiparty computation (MPC) committee}. We already presented how the selection of the committee process is done in the assumptions section \ref{sec:assumptions}. These committee members, each holding only partial values of \( x \) and \( r \), collectively sum the balances in an MPC manner. Concretely, if the LPs balances are \( \{x_1, x_2, \dots, x_n\} \), the committee runs an off-chain MPC addition protocol to compute:
\begin{equation}
     X \;=\; \sum_{i=1,j=1}^{n,l} x_{i_j}
\end{equation}
without revealing the individual \( x_i \), where $n$ is the number of LPs and $l$ is the number of shares for each LP.
. At the conclusion of this phase, only the \emph{aggregate liquidity} \( X \) will be revealed for price discovery. so each individual’s contribution remains confidential, yet the system can publicly confirm that \( \sum_{i,j} x_{i,j} \) is correctly computed. \\
\textbf{2) Trading Phase:}
Once the total liquidity is disclosed, \emph{trading} proceeds via private buy or sell orders, ensuring front-running and sandwich attacks are mitigated. A trader with balance \( y \) wanting to buy (or sell) an asset again forms a Pedersen commitment \( \mathrm{Comm}(y, r') \) and supplies a zero-knowledge proof attesting to ownership of \( y \). They then privately send their blinding factor \( r' \) (and secret-shares of \( y \)) to the MPC committee. The committee executes the Beaver-triple multiplication to compute the AMM formula. for instance, in a \emph{constant-product} market maker, we have: 
\begin{equation}
    E.  M =  C
\end{equation}
so a buyer spending \( \Delta M = m\) to acquire \( \Delta E = e \) would satisfy:
{
\begin{multline}   
    C \;=\; (M+m)(E+e) \\
     \;=\; \\
     \underbrace{m\,e}_{\text{needs one Beaver triple}}
     \;+\; M\,e
     \;+\; E\,m
     \;+\; M\,E .
\end{multline}
We need only one Beaver triple to obtain a secret sharing of $m.e$. The other three terms need no interaction:
\begin{itemize}
    \item Multiplying a public constant by a secret $M.e$ and $E.m$ is just a local scaling of every share.
    \item $M.E$ is a known constant that can be added in plaintext or secret-shared with a single random mask.
\end{itemize}
In the offline phase, the committee jointly generates the Beaver triple:
\begin{equation}
\bigl({a},{b},{c}\bigr),
\qquad
c = a\,b .
\end{equation}
Then, each party $P_i$ broadcasts $(d_i,e'_i)$ and reconstructs the masks:
\begin{align}
d_i      &= m_i - a_i , &
e'_i     &= e_i - b_i      \\
d        &= \sum_{i=1}^n d_i , &
e'       &= \sum_{i=1}^n e'_i \\
{z_i} &= {c_i}
            + d\,{b_i}
            + e'\,{a_i}
            + \frac{d\,e'}{n} &&
            \text{(share of } m\,e \text{)}
\end{align}
Finally, Each party computes its final share of $C$:
\begin{equation} 
{C_i}
   = {z_i}
   + M\,{e_i}
   + E\,{m_i}
   + \delta_{i1}\,M\,E ,
\end{equation}

where $\delta_{i1}$ is the Kronecker delta (only party 1 keeps the public constant $ME$ so the vector sums correctly). \\
Together $\bigl({C_1},\dots,{C_n}\bigr)$ form a valid sharing of the desired:
\begin{equation}
C \;=\; (M+\sum_{i=1}^n M_i)\,(E+\sum_{i=1}^n E_i).
\end{equation}
The multiplication completes in just one broadcast round, the minimum for an actively secure Beaver-style MPC gate. ensuring the privacy of inputs and securing users from maximum extractable value (MEV) attacks.

\begin{figure*}
    \centering
    \includegraphics[scale=0.12]{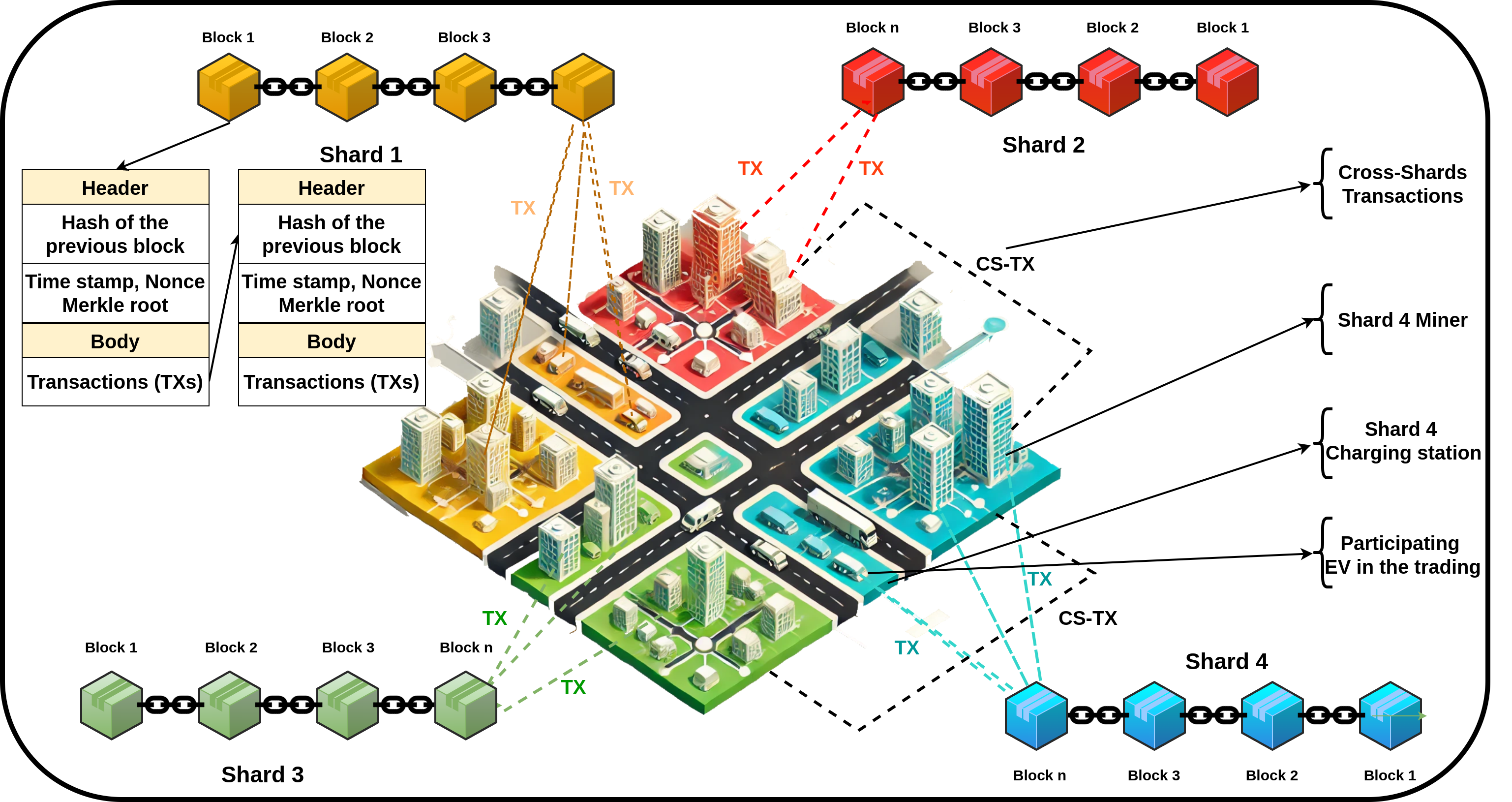}
    \caption{Deployment Architecture }
    \label{fig:sharding}
\end{figure*}

\section{Cryptographic Building Blocks}
\label{sec:bblocks} The main cryptographic building blocks upon which the $EnerSwap$ system is built are the following:

 \quad \textbf{Pedersen commitment.} 
  A cryptographic technique that enables a user to commit to a message $m$ without disclosing it to the receiver \cite{pedersen1991non}. The security of a commitment scheme relies on two properties: 
 \begin{itemize}  
   \item \textit{Hiding:} Given $\mathsf{COMM(x)}$, it should be computationally infeasible to determine the original value $x$.
   \item \textit{Binding:} It should be computationally infeasible to find two distinct values $x_1$ and $x_2$ s.t. $\mathsf{COMM}(x_1) = \mathsf{COMM}(x_2)$.
 \end{itemize}  

\quad \textbf{zkSNARKs.} \textit{Zero-Knowledge Succinct Non-Interactive Arguments of Knowledge} are a type of non-interactive zero-knowledge proof (NIZK) that enables a prover to convince a verifier of a statement’s validity without revealing any information beyond its correctness. zkSNARKs offer key properties such as zero-knowledge, soundness, completeness, succinctness, and non-interactivity \cite{groth, plonk}.

\quad A zkSNARK (e.g., Groth16) comprises four main algorithms \cite{groth}: 
\begin{itemize} \item $\mathsf{Setup}(1^\lambda)$: Generates public parameters $pp$ from a security parameter $\lambda$. \item $\mathsf{Gen}(pp, C)$: Produces a proving key $pk$ and a verification key $vk$ for a computation $C$. \item $\mathsf{Prov}(pk, t, w)$: Generates a proof $\pi$ that the prover knows a private witness $w$ satisfying the computation $C$ on public input $t$. \item $\mathsf{Verif}(vk, t, \pi)$: Verifies the proof $\pi$ against input $t$ and returns true if valid.
\end{itemize}

\subsection{Proof construction}
\begin{algorithm}[t]
\footnotesize
\caption{Balance-at-Least zkSNARK}

\KwIn{
  \textbf{Private:}  \(\texttt{leafRlp} \in \{0,\dots,255\}^{112}\)\;
  \textbf{Public:}   \(\texttt{k\_hi}, \texttt{k\_lo} \in [0,2^{128}-1]\),\;
                     \(\texttt{leafHash} \in \mathbb{F}^{4}\)}
\KwOut{\(\texttt{proof}\) — zkSNARK attesting “balance \(\ge k\)”}

\BlankLine
\(\texttt{h\_check} \leftarrow \mathrm{Keccak256}(\texttt{leafRlp})\)\;
\textbf{assert} \(\texttt{h\_check} = \texttt{leafHash}\)\tcp*{leaf integrity}

\BlankLine
\((\texttt{bal\_hi},\texttt{bal\_lo}) \leftarrow \textsf{decodeBalance}(\texttt{leafRlp})\)\;
\tcp{\small reads 32-byte balance starting at byte-offset 3}

\BlankLine
\(\texttt{sufficient} \leftarrow
   (\texttt{bal\_hi} > \texttt{k\_hi})
   \;\lor\;
   (\texttt{bal\_hi} = \texttt{k\_hi} \land \texttt{bal\_lo} \ge \texttt{k\_lo})\)\;

\textbf{assert} \(\texttt{sufficient}\)\tcp*{balance meets threshold}

\BlankLine
\(\texttt{proof}   \leftarrow \textsf{zkSNARK.Prove}\bigl(
      \texttt{circuit},
      \texttt{private\_inputs}= \{\texttt{leafRlp}\},
      \texttt{public\_inputs} = \{\texttt{k\_hi},\texttt{k\_lo},\texttt{leafHash}\}\bigr)\)\;

\Return \((\texttt{proof}, \{\texttt{k\_hi}, \texttt{k\_lo}, \texttt{leafHash}\})\)\;
\label{algo:balanceAtLeastPseudo}
\end{algorithm}
We now explain how zkSNARKs are integrated into $EnerSwap$, outlining the system setup, participating roles, and proof generation process. \\
\quad \textit{{\textbf{ 1-Setup Phase:}}}  zkSNARK systems typically require an initial trusted setup phase, commonly known as the \textit{Powers-of-Tau} ceremony. This phase generates structured public parameters, collectively referred to as the \textit{Common Reference String} (CRS)\cite{zkpCrs}, or $pp$, which are used to derive the proving key ($pk$) and verification key ($vk$). These keys are essential for proof generation and verification, respectively, ensuring that both parties operate over the same cryptographic structure tied to the target computation. The setup typically involves multiple independent participants contributing entropy, which collectively ensures that no single entity can compromise the security of the system. In $EnerSwap$, this setup is performed once, independently of any specific trading activity.

\paragraph*{Roles in the Protocol} $EnerSwap$ defines two primary roles in the zkSNARK-based billing verification process:

\begin{itemize} \item \textbf{Prover (Trader or Liquidity provider):} The trader constructs a zkSNARK proof demonstrating that their account balance lies within a permitted range. The proof is computed over private data $LeafRLP$ (the account’s confidential state) and public data (LeafHash, and the interval of the balance wants to prove $K_{hi}$ and  $K_{lo}$) that allows him to proceed with the trade. Section \ref{sec:protocol} explains the Merkle Patricia Trie (MPT) proof mechanics and shows how the tree’s structure enables balance verification in a zero-knowledge-proof (ZKP) framework.
\item \textbf{Verifier (Smart Contract and External Parties):} Verification is mainly handled on-chain by a smart contract that automatically checks the validity of submitted proofs using the verification key $vk$. Because zkSNARKs are publicly verifiable, any external entity with access to $vk$ and the public inputs can also verify the proof, without accessing the trader’s private data. 
\end{itemize}

\textbf{\textit{{2- Proof Generation}}} After setup, the traders and Liquidity providers who want to prove certain liquidity or balance execute the proof construction process detailed in Algorithm \ref{algo:balanceAtLeastPseudo}.
This process involves evaluating an arithmetic circuit that encodes the MPT logic: It verifies that the constructed RLP-encoded leaf (LeafRLP) and its hash (LeafHash) are both consistent with the hash up to the provided state root. The proof $\pi$ produced by this circuit attests to the freshness and Unforgeability of the balance while preserving the privacy of the private data.

\section{Deployment Architecture}
Deploying such a solution in a distributed manner requires an architecture that can coordinate a global marketplace yet still respect the physical realities of power grids and the patchwork of national regulations that govern them. Our design, therefore, adopts a two-tier blockchain architecture as illustrated in Figure \ref{fig:sharding}:
\begin{itemize}
    \item \textbf{The masterchain} acts as the protocol’s backbone. It stores global parameters, validator sets and their stakes, a registry of all active workchains and their shards, and—most critically, the hash of the latest block from every shardchain. By anchoring these state roots, the masterchain gives the entire system a single source of truth for cross-chain finality, slashing conditions, and governance upgrades, while remaining lightweight enough to reach consensus quickly.
    \item  \textbf{Workchains} are the workhorses. Each corresponds to a distinct geographic or regulatory zone and handles the actual value transfer and smart contract transactions, such as bids, clears, and energy certificate settlements. Because different regions may require different address formats, transaction types, virtual machines, or even native tokens, each workchain is free to customize its rules. Interoperability is maintained through a set of mandatory bridging primitives and cryptographic proofs enforced by the masterchain, allowing assets and messages to move atomically across zones.
\end{itemize}
Within every workchain, dynamic geographical sharding further splits the state when local load spikes and merges shards when demand subsides, ensuring low-latency clearing without over-provisioning resources. Local consensus remains inside the shard, keeping settlement times in the millisecond-to-second range, while periodic checkpoints flow upward to the masterchain for global finality. If a fault, say, a hurricane in Texas, takes validators offline, disruption is contained to the affected shard; other regions' markets continue operating uninterrupted.
\section{Security Analysis}
\label{sec:security}
\quad In this section, we assess the security guarantees offered by the EnerSwap framework in light of the threat model presented in Section ~\ref{sec:security-prop}. We first start with the attacks on the blockchain system, then we discuss the trading attacks\\
\subsection{Blockchain Attacks}
\textbf{ 1) Sybil Attack.}
We consider that EnerSwap operates over a public or consortium blockchain platform that inherently is Sybil-resistant, consisting of a fixed set of ${n = 3f + 1}$ replicas, indexed by i $ \in [n]$ where $[n] = {1, . . . , n}$. A set $F \subset [n]$ of up to $f = |F |$ replicas are Byzantine faulty, and the remaining ones are correct. An adversary $\mathcal{A}$, can learn all internal states held by these replicas, including their public cryptographic keys.
The underlying consensus protocol guarantees safety and liveness provided that at most $f$ replicas are faulty and the network is eventually synchronous.
Hence, the ledger offers a totally ordered, immutable log that $\mathcal{A}$ cannot fork or reorder once a block is final. \\
The rest of the components of the framework, including ZKP generation and user commitments, do not rely on the assumption of a majority honest participant base. Furthermore, actors such as LPs and Traders are uniquely identified by their cryptographic keys and smart contract-bound identities, preventing identity duplication. As a result, even if an adversary $\mathcal{A}$ creates multiple Sybil identities, they gain no additional influence over the outcome of billing settlements. \\
\textbf{Resistance to Poisoning Attacks:}  EnerSwap is explicitly designed to eliminate trust assumptions regarding the honesty of the traders in reporting their balance account and their ability to trade. EnerSwap removes this assumption by requiring buyers and sellers to generate a ZKP that demonstrates:
    \begin{enumerate}
        \item The actual account balance amount by computing the stateroot and leafhash
        \item The reported balance used in the MPC is hashed into shares previously committed by the trader and stored in the blockchain.
    \end{enumerate}
    This mechanism ensures that the traders cannot manipulate the committee to accept trades that are bigger than their account balance, which are verified on-chain by the deterministic contract $\mathtt{Verify}$  deployed with verifying key $\mathsf{vk}$. Safety of the $3f\!+\!1$ BFT ledger ensures all honest replicas execute identical bytecode, so a proof is accepted globally if
    \begin{equation}
        f( \mathtt{Verify}(\pi,x))=1
    \end{equation}
   Consequently, knowledge-soundness and zero-knowledge reduce to Discrete-Log Hardness in two prime-order groups assumptions (Groth16 assumption); For this,  no extra honest-majority requirement arises beyond the $f<\tfrac{n}{3}$ bound already necessary for consensus. this architecture robustly defends against poisoning attacks. \\
\textbf{3) Resistance to On-Chain Privacy Leakage.} EnerSWAP addresses this issue through:
    \begin{definition}[Hiding]
    \label{def:hide}
    Given $n$ equal-length messages $m_0,m_1,...,m_n, \mathcal{A}'s $ advantage in guessing which one was committed is negligible.
    The \emph{hiding advantage} of $\mathcal A$ is defined as
    \begin{equation}
        \operatorname{Adv}^{\mathrm{hide}}_{\mathcal A}(\lambda)
        \;:=\;
        \bigl|\Pr[\textsf{win}]-\tfrac{n}2\bigr|.
    \end{equation}
    \end{definition}

    \begin{definition}[Computational Zero-Knowledge]\label{def:zk}
    $\mathcal{A}$’s view  of a published proof is computationally indistinguishable from a simulator $S$ that does not know the witness. where $S$ is a hypothetical, efficient algorithm that given only the public statement can output a transcript indistinguishable from what any real verifier would see, thereby proving that the verifier gains no knowledge of the secret witness during the actual protocol.

for every probabilistic polynomial-time (PPT) verifier $V^{\ast}$  
there exists a PPT simulator $S$ such that, for all security parameters 
$\lambda$ and all $(x,w)\in R$, the ensembles
\begin{equation}   
  \bigl\{
      \operatorname{View}_{V^{\ast}}^{\,P(x,w)}(\lambda)
  \bigr\}_{\lambda,x,w}
  \quad\text{and}\quad
  \bigl\{
      S^{V^{\ast}}(x)(\lambda)
  \bigr\}_{\lambda,x,w}
\end{equation}

are computationally indistinguishable.
\\
For every PPT distinguisher $\mathcal D$,
\begin{equation}
\begin{aligned}
  \bigl|
    \Pr\!\bigl[
      \mathcal D\bigl(
        \operatorname{View}_{V^{\ast}}^{P(x,w)}(\lambda)
      \bigr)=1
    \bigr]
    -
\\[2pt]
    \Pr\!\bigl[
      \mathcal D\bigl(
        S^{V^{\ast}}(x)(\lambda)
      \bigr)=1
    \bigr]
  \bigr|
  \;\le\;
  \operatorname{negl}(\lambda).
\end{aligned}
\end{equation}

Here, $\operatorname{View}_{V^{\ast}}^{\,P(x,w)}(\lambda)$ denotes everything
$V^{\ast}$ sees during (or as output of) its interaction with~$P$,
while $S^{V^{\ast}}(x)(\lambda)$ is the simulator’s output generated
\emph{without} knowledge of the witness~$w$.
\end{definition}

No actual usage data is ever submitted on-chain. Instead, the traders and LPs compute the PCs of their shares, which serve as a non-reversible commitment to the Committee. The zero-knowledge proof guarantees that the revealed total corresponds to the private account balance that matches the PC, without revealing the actual values. This design prevents adversaries from inferring sensitive details about a user's trading behavior by monitoring blockchain transactions, thereby ensuring strong privacy even in open networks.
\subsection{Trading Attacks}
\quad In this section, we analyze the proposed scheme's potential security vulnerabilities, specifically focusing on trading-related attacks. We also demonstrate the scheme's resilience against these threats. It is important to highlight that this security analysis is informal, initially discussing possible threats. An extended version of this work will provide a more comprehensive and detailed analysis.

Our protocol’s primary defense against malicious MEV (Maximal Extractable Value) activities and trading attacks stems from its confidentiality guarantees, formally proven in the section before, during both the initialization and trading phases. In standard AMM implementations, adversaries can observe pending transactions in the mempool and strategically reorder or front-run them, extracting value from large or time-sensitive trades. By contrast, this protocol conceals individual trade sizes and defers the public revelation of key pool states until after trades finalize. 
Adversaries cannot reliably detect the magnitude or timing of a forthcoming trade, thus thwarting typical front-running and sandwich tactics. Specifically, because each buy or sell operation is $committed$ via a $Pedersen scheme$ on-chain and processed off-chain via MPC, any public indication of the trade comes only from a final aggregated pool update. Even then, adversaries can only observe the updated total or price, rather than the raw transaction size or pre-confirmation details. 

From a formal standpoint, assume an adversary has full visibility into network mempool contents aside from committed transactions, where they see the commitments but cannot extract the hidden amounts. The adversary learns no useful information about the trader’s exact balance or intended trade size from the published commitments. Likewise, the off-chain MPC phase guarantees correctness of arithmetic operations such as maintaining \ref{eq:AMM} in a constant-product, without revealing partial sums or intermediate values. Thus, adversarial participants or external observers lack the critical data needed to place sandwiching orders around the victim’s transaction. This design significantly reduces the scope of MEV manipulations (front-running, back-running, value extraction from timing attacks) and preserves fair market conditions for liquidity providers and traders alike. Consequently, attacks that rely on publicly visible transaction details, such as sandwiched trades or sniping events, are rendered computationally infeasible or unprofitable, thereby enhancing the overall security and fairness of the trading environment.
\section{Implementation, Evaluation and Results}
\label{sec:eva}
\subsection{Implementation and Experimental setup}
\quad All arithmetic circuits were written in Circom\footnote{\href{https://github.com/iden3/circom}{https://github.com/iden3/circom}}, while the on-chain verification logic supporting Groth16 \cite{groth}, Plonk \cite{plonk} backends was implemented in Solidity and deployed to a private 4-node Geth network; commitment primitives were developed in Python, and the MPC setup was executed and evaluated with SPDZ \cite{SPDZ}. Performance was evaluated on a Dell XPS 15 9530 laptop (Intel Core i7-12700H, 32 GB RAM) using Hyperledger Caliper\footnote{\href{https://github.com/hyperledger/caliper-benchmarks}{https://github.com/hyperledger/caliper-benchmarks}}, we measured proof-generation time, verification gas cost, latency, and throughput; Groth16 and Plonk proofs were produced with SnarkJS\footnote{\href{https://github.com/iden3/snarkjs}{https://github.com/iden3/snarkjs}}. \\
Because the TON \cite{ton2021whitepaper}  Virtual Machine used for dynamic sharding deployment differs from the EVM, we also executed a simulation contract in the TON network.
The full code of the paper is available on Github \footnote{\href{https://github.com/0xmoncif213/EnerSwap}{https://github.com/0xmoncif213/EnerSwap}}.
\subsection{Performance Evaluation}
\quad We consider three metrics for $EnerSwap$ performance evaluation as described below:
\begin{itemize}[leftmargin=0.2cm,align=left]
\item \textbf{Throughput:} Number of successful transactions (TXs) per second (tx/s).
\item \textbf{Latency:} Time interval between TXs submission and its validation on-chain.
\item \textbf{Gas:} unit that measures the computational work required to perform operations.
\end{itemize}
\subsection{Results discussion}

\begin{table}[th]
\centering
\caption{Proving and verification overhead results.}
\label{tab:proof-results}
\begin{tabular}{|l|c|c|}
\toprule
\textbf{Proof} & \textbf{Groth16} & \textbf{Plonk} \\ 
\midrule
Generation time ($_{S}$)  & 0.9  & 2.11   \\ 
Proof size ($_{B}$)  & 805  & 2300   \\ 
Generation memory consumption ($_{KB}$)  & 143512            & 205036   \\ 
Verfication gas consumption ($_{GAS}$) & 218000          & 260350  \\ 
\bottomrule
\end{tabular}
\end{table}
\quad The evaluation is divided into three main parts:
\begin{itemize}
    \item \textit{Simulation Results:}
\end{itemize}
We executed identical contracts in term of complexity on both TON and Ethereum for three successive blocks, and the gas costs on Ethereum-PoA and the TON work-chain were almost identical, indicating the same computational effort. However, latency diverged sharply: TON remained flat at about $4.5 {s}$, whereas Ethereum-PoA grew from $4 {s}$ to $5 {s}$ and finally $11 s$ as the block gas doubled and then quadrupled. This underscores TON’s ability to scale better under high-frequency environments. Because TON is still maturing and its sharding model needs further adaptation for our use case. A prototype based on a modified TON workchain and its full deployment will be detailed in the extended version of this work. 
\begin{itemize}
\item \textit{Off-chain overhead:} including proof generation time, memory usage, and MPC Framework offline and online latency.
\end{itemize}
Table \ref{tab:proof-results} shows that Groth16 outperforms Plonk across three critical dimensions: it generates proofs in just $0.9_{s}$ versus$ 2.1_{s}$, consumes only $144_{MB}$ of RAM compared with $205_{MB}$, and produces a far more compact proof $805_{B}$ against Plonk’s $2.3_{kB}$, making Groth16 the clear choice for latency-sensitive, resource-constrained deployments or off-chain clients, although this advantage must still be weighed against the on-chain verification. \\
Figure \ref{fig:offline-vs-online}  breaks down protocol offline and online latency; we deliberately omit end-to-end latency because real deployments are affected by how long participants take to supply their inputs, a delay that is external to the protocol and would distort any analysis. The first three bars report the Init setup phase (offline) for each peer, as the task is essentially a single secure addition. The red bars chart the online phase, where the MPC engine processes live trades. Online latency climbs with the number of peers because every extra player adds multiplication gates and an additional communication round, yet the growth is modest: $7_{ms}$ with three peers and only  $33_{ms}$ with ten. Trading phase still costs more than init phase, multiplications are inherently heavier than additions, but the absolute numbers remain comfortably below the sub $1_{s}$ threshold usually cited \cite{icbc}. For real-time interactions, confirming the protocol scales gracefully even as the pool of liquidity providers and traders expands.
\begin{itemize}
\begin{figure*}[t]     
    \centering
    \subfloat[Latency and Throughput]{
        \includegraphics[scale=0.32]{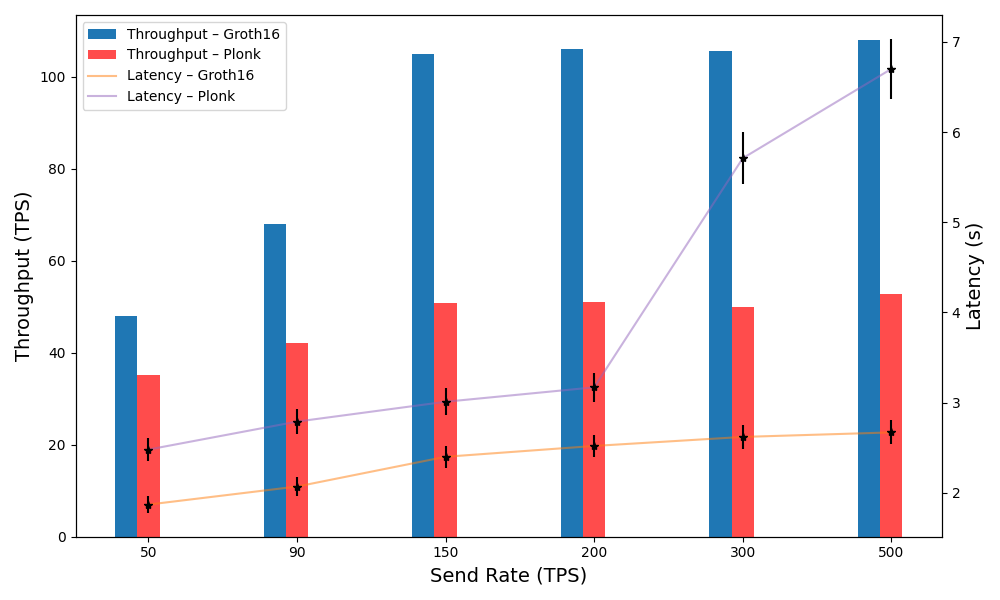}
    \label{fig:throughput}
   }
    \subfloat[On-chain Gas Consmuption ]{
         \includegraphics[scale=0.32]{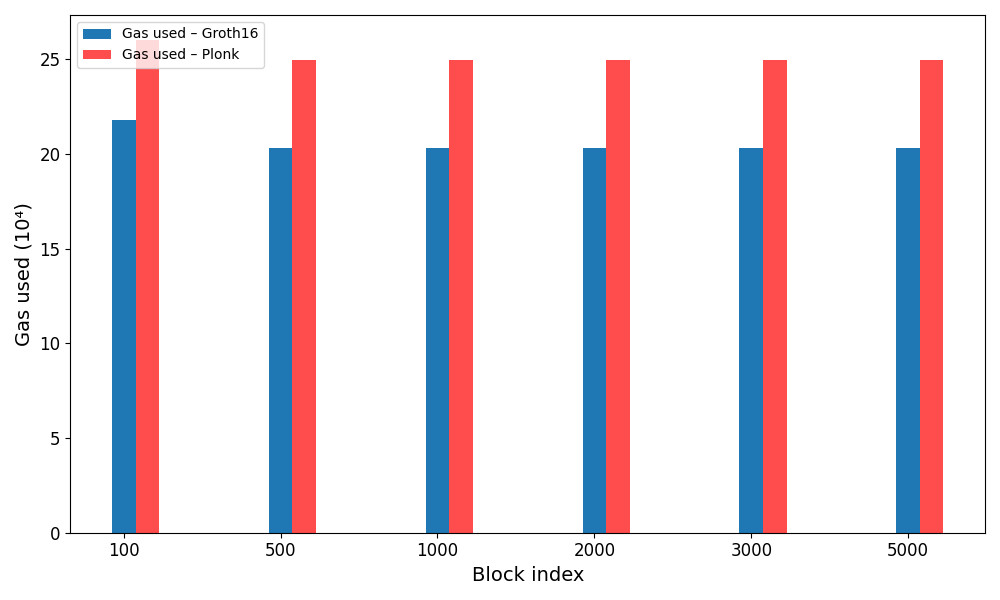}
        \label{fig:gas}
   }
  
    \caption{ \centering On-chain Performance : A Comparison Between Groth16 and Plonk implementations}
    \label{createTaskRep-gas}
\end{figure*}
\item \textit{On-chain verification overhead:} including gas consumption, latency, and throughput.
\end{itemize}
We now turn to the blockchain network performance. \ref{fig:gas} compares the gas required for the on-chain verification routines implemented with Groth16 and Plonk. For both proving systems, the cost remains essentially flat as block height grows; during a 24-hour soak test that mined more than 3,000 blocks, We saw a small bump in gas usage on the very first calls; expected, since they allocate new storage, but after that the cost drops slightly and then stays flat. Identical computations consistently incur identical fees, underscoring the implementation’s stability and its suitability for long-running deployments.
Figure \ref{fig:throughput} puts both verifier contracts through a closed-loop stress test that steadily drives up the transaction send rate. Groth16 tops out at just over $100_{TPS}$, whereas Plonk plateaus around $55_{TPS}$, confirming Groth16’s higher sustainable throughput. Latency naturally increases as the send rate grows, more pending transactions deepen the mempool, but Groth16’s response time stays almost flat because its on-chain verifier has $O(1)$ complexity, so each additional proof adds a constant, tiny verification cost. Plonk, by contrast, shows a steeper latency curve, reflecting its heavier per-transaction workload. The result is a clear throughput–latency advantage for Groth16 under high-load conditions.
Collectively, these findings confirm that $EnerSwap$ using Groth16 not only preserves data privacy but also unlocks the transaction volume required for real-world deployments.
\begin{figure}
    \centering
    \includegraphics[scale=0.34]{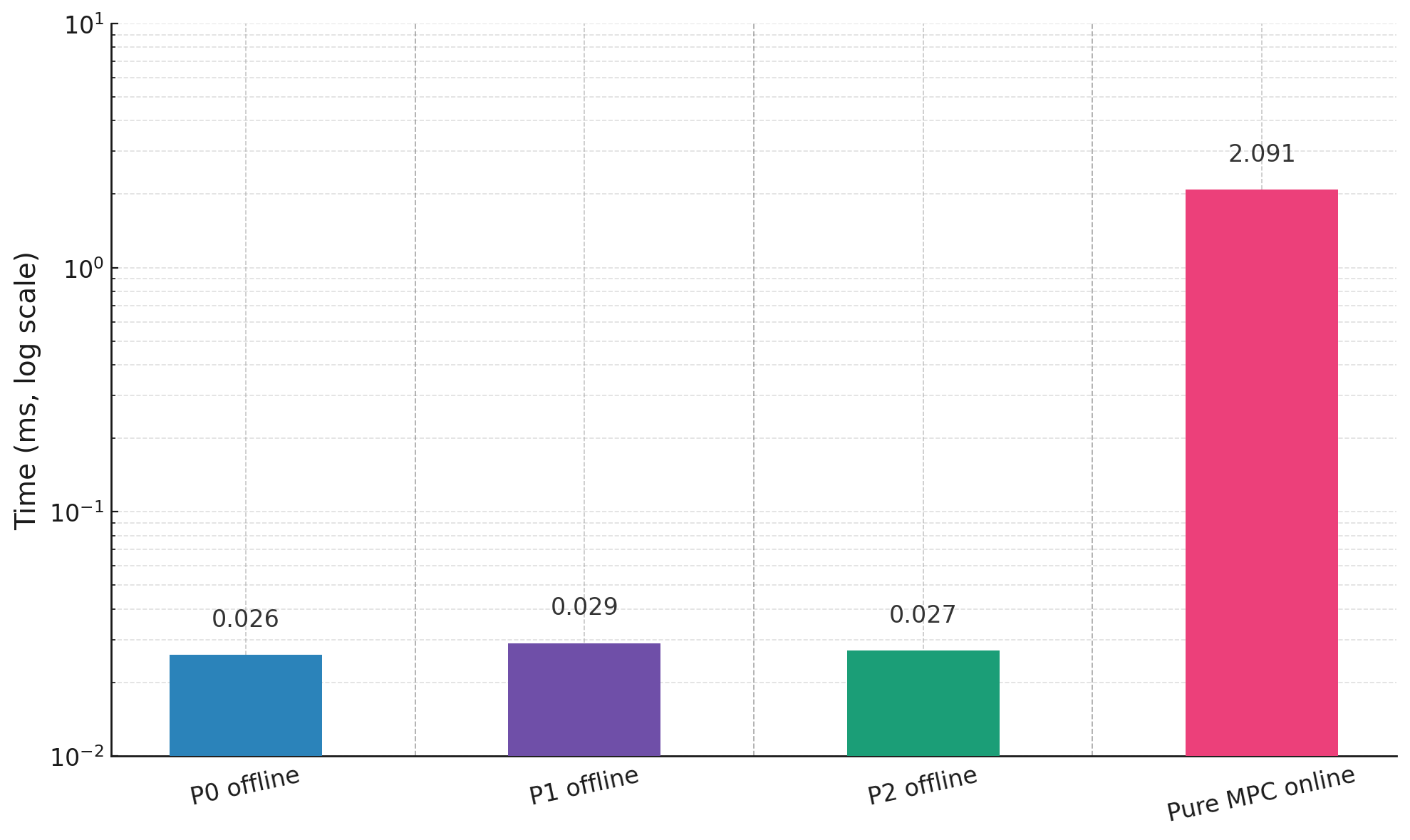}
    \caption{MPC-init-Phase: Offline VS Online Latency with 3 players}
    \label{fig:offline-vs-online}
\end{figure}
\begin{figure}
    \centering
    \includegraphics[scale=0.34]{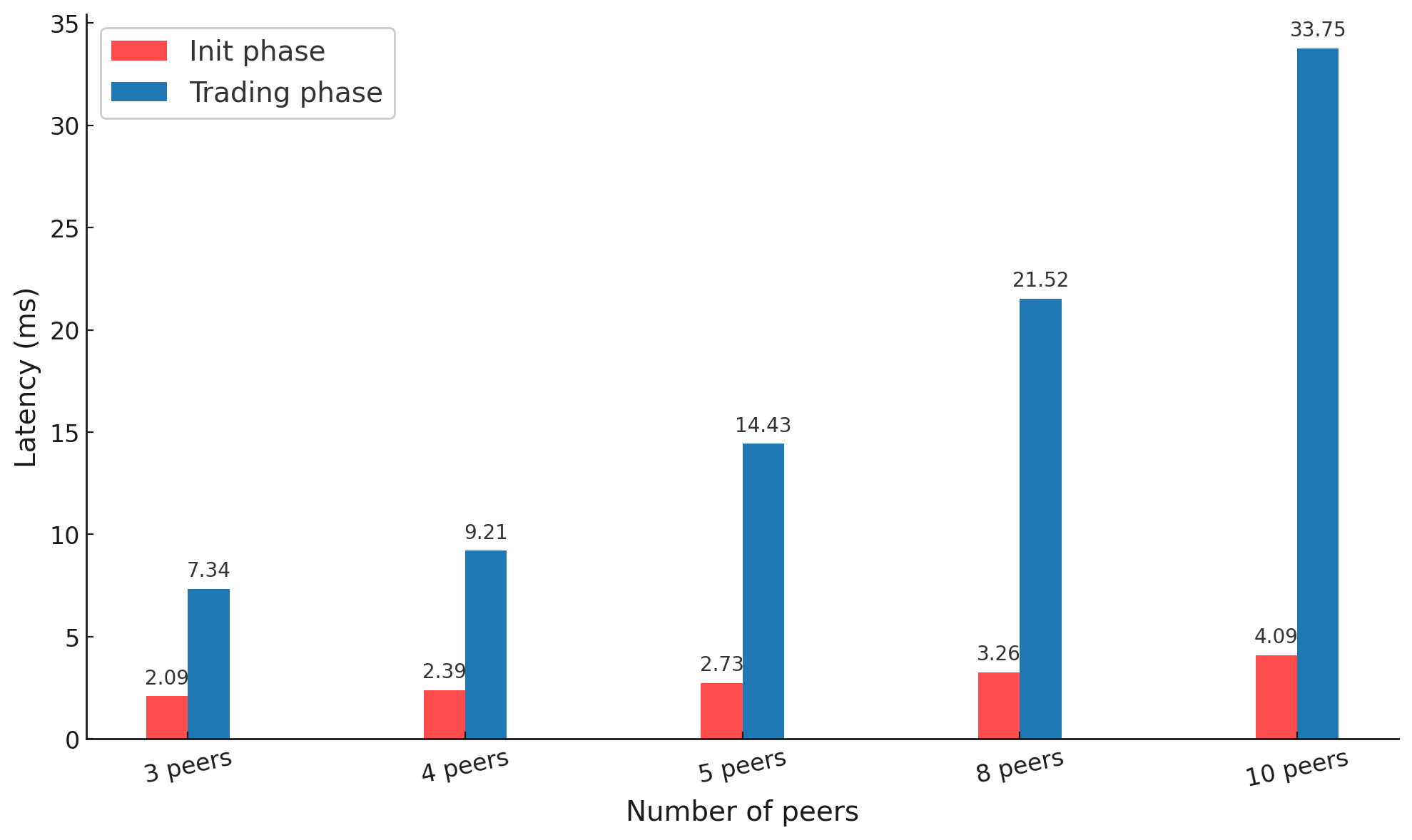}
    \caption{MPC-init vs trading: Online Latency with multiple players}
    \label{fig:offline-vs-online}
\end{figure}

\section{Conclusion}
\label{sec:conclusion}
This paper introduces EnerSwap, a fully decentralized, privacy-preserving AMM for energy trading. EnerSwap uses geographically sharded blockchain architecture and multi-party computation (MPC) with ZKPs and commitment schemes to process trades off-chain and posts succinct proofs on-chain to certify correctness, while a uniform clearing-price algorithm guarantees price fairness and protects order flow and account-balance confidentiality. Our benchmarks show that the current design scales to real-world trading volumes. For future work, we outline two key directions: privacy-preserving concentrated-liquidity pools further to deepen market liquidity, and role-specific incentive mechanisms to align the interests of liquidity providers, traders, and verifiers and ensure long-term ecosystem health.


\section*{Acknowledgment}
This work is supported by the OPEVA project, which has received funding within the Chips Joint Undertaking (Chips JU) from the EU’s Horizon Europe Programme and the National Authorities (France, Czechia, Italy, Portugal, Turkey, Switzerland), under grant agreement 101097267. In France, the project is funded by BPI France under the France 2030 program on ``Embedded AI''. Views and opinions expressed are, however, those of the authors only and do not necessarily reflect those of the EU or Chips JU. Neither the EU nor the granting authority can be held responsible for them.

\end{document}